\documentstyle [aps,multicol,psfig]{revtex}
\renewcommand{\narrowtext}{\begin{multicols}{2} \global\columnwidth20.5pc}
\renewcommand{\widetext}{\end{multicols} \global\columnwidth42.5pc}

\begin{document}

\newcommand{\newc}{\newcommand}

\newc{\be}{\begin{equation}}
\newc{\ee}{\end{equation}}
\newc{\ba}{\begin{eqnarray}}
\newc{\ea}{\end{eqnarray}}
\newc{\bea}{\begin{eqnarray*}}
\newc{\eea}{\end{eqnarray*}}
\newc{\D}{\partial}
\newc{\ie}{{\it i.e.} }
\newc{\eg}{{\it e.g.} }
\newc{\etc}{{\it etc.} }
\newc{\etal}{{\it et al.}}
\newcommand{\nn}{\nonumber}

\newc{\ra}{\rightarrow}
\newc{\lra}{\leftrightarrow}
\newc{\no}{Nielsen-Olesen }
\newc{\lsim}{\buildrel{<}\over{\sim}}
\newc{\gsim}{\buildrel{>}\over{\sim}}

\title{Q Rings}
\author{M. Axenides$^a$, E. Floratos$^a$, S. Komineas$^b$, L. Perivolaropoulos$^a$,
}
\address{$^a$ Institute of Nuclear Physics, N.C.R.P.S. Demokritos,  153
10, Athens, Greece \\ $^b$ Physikalisches Institut, Universit\"at 
Bayreuth, D-95440 Bayreuth, Germany.} 

\date{\today}
\maketitle

\begin{abstract}

We show the existence of new stable ring-like localized scalar 
field configurations whose stability is due to a combination of 
topological and nontopological charges. In that sense these 
defects may be called semitopological. These rings are Noether 
charged and also carry Noether current (they are superconducting). 
They are local minima of the energy in scalar field theories with 
an unbroken U(1) global symmetry. We obtain numerical solutions of 
the field configuration corresponding to large rings and derive 
virial theorems demonstrating their stability. We also derive the 
minimum energy field configurations in 3D and simulate the 
evolution of a finite size Q ring on a three dimensional lattice 
thus generalizing our demonstration of stability. 

\end{abstract}

\narrowtext
 Non-topological solitons (Q balls) have been studied 
extensively in the literature in one, two and three 
dimensions\cite{lp92}. They are localized time dependent field 
configurations with a rotating internal phase and their stability 
is due to the conservation of a Noether charge $Q$\cite{c85}. In 
three dimensions, the only localized, stable configurations of 
this type have been assumed to be of spherical symmetry hence the 
name Q balls. The generalization of two dimensional (planar) Q 
balls to three dimensional Q strings leads to loops which are 
unstable due to tension. Closed strings of this type are naturally 
produced during the collisions of spherical Q balls and have been 
seen to be unstable towards collapse due to their 
tension\cite{bs00,lpwww}. 

There is a simple mechanism however that can stabilize these 
closed loops. It is based on the introduction of an additional 
phase on the scalar field that twists by $2\pi N$ as the length of 
the loop is scanned. This phase introduces additional pressure 
terms in the energy that can balance the tension and lead to a 
stabilized configuration, the {\it Q ring}. This type of pressure 
is analogous to the pressure of the superconducting string 
loops\cite{Witten:1985eb} (also called `springs'\cite{hht88}). In 
fact it will be shown that Q rings carry both Noether charge and 
Noether current and in that sense they are also superconducting. 
However they also differ in many ways from superconducting 
strings. Q rings do not carry two topological invariants like 
superconducting strings but only one: the winding $N$ of the phase 
along the Q ring. Their stability is due to the conservation of 
both the topological twist and the nontopological Noether charge. 
Hence Q rings may be viewed as semitopological defects. In 
contrast to Q balls, they are local minima of the Hamiltonian 
separated from Q balls by a finite energy barrier. In what follows 
we demonstrate the existence and metastability of Q rings in the 
context of a simple model. We use the term 'metastability' instead 
of `stability' because {\it finite size} fluctuations can lead to 
violation of cylindrical symmetry and decay of a Q ring to one or 
more Q ball as demonstrated by our numerical simulations. 

Consider a complex scalar field whose dynamics is determined by 
the Lagrangian \be \label{model} {\cal L}={1\over 2} 
\partial_\mu \Phi^* 
\partial^\mu \Phi - U(|\Phi |) \ee
The model has a global $U(1)$ symmetry and the associated
Noether current is \be \label{current} J_\mu = Im(\Phi^* 
\partial_\mu \Phi) \ee with conserved Noether charge $ Q=\int 
d^3 x \; J_0 $.  Provided that the potential of (\ref{model}) 
satisfies certain conditions \cite{c85,lp92} the model accepts 
stable Q ball solutions which are described by the ansatz $ \Phi = 
f(r) e^{i \omega t}$.  The energy density of this Q ball 
configuration is localized and spherically symmetric. The 
stability is due to the conserved charge $Q$. 

In addition to the $Q$ ball there are other similar stable 
configurations with cylindrical or planar symmetry but infinite, 
not localized energy in three dimensions. For example an infinite 
stable Q string that extends along the z axis is described by the 
ansatz \be \label{stranz} \Phi = f(\rho) e^{i \omega t}\ee where 
$\rho$ is the azimouthal radius.  This configuration has also been 
called `planar' or 'two dimensional' Q ball\cite{lp92}. 

The energy of this configuration can be made finite and localized 
in three dimensions by considering closed Q strings. These 
configurations which have been shown to be produced during 
spherical Q ball collisions\cite{bs00,lpwww} are unstable towards 
collapse due to their tension. In order to stabilize them we need 
a pressure term that will balance the effects of tension. This 
term appears if we substitute the string ansatz (\ref{stranz}) by 
the ansatz of the form \be \label{qringanz}\Phi = f(\rho) e^{i 
\omega t} e^{i \alpha(z)} \ee where $\alpha(z)$ is a phase that 
varies uniformly along the z axis. This phase introduces a new 
non-zero $J_z$ component to the conserved current density 
(\ref{current}). The corresponding current is of the form \be 
\label{jzcons} I_z=  \int d z \;{{d \alpha}\over {d z}} \; 
2\pi\int d\rho \; \rho \; f^2 \ee  Consider now closing the 
infinite Q string ansatz (\ref{qringanz}) to a finite (but large) 
loop of size $L$. The energy of this configuration may be 
approximated by \bea E &=& {{Q^2}\over {4 \pi L \int d \rho \; 
\rho \; f^2}} + \pi \; L \; \int d \rho \; \rho \; f'^2 \nn 
\\ &+& {{(2\pi N)^2 \pi}\over L}\int d \rho \; \rho \; f^2 + 2 \pi  L \int d 
\rho \; \rho U(f)\nn \\ &\equiv & I_1 + I_2 + I_3 + I_4 \eea where 
we have assumed $\alpha (z) = {{2 \pi N}\over L} z $ and the terms 
$I_i$ are all positive. Also $Q$ is the charge conserved in $3D$ 
defined as 
\be
Q=\omega 2\pi L\int d\rho \; \rho \;f^2 \ee The winding $2 \pi 
N=\int d z \;{{d \alpha}\over {d z}}$ is topologically conserved 
and therefore the current (\ref{jzcons}) is very similar to the 
current of superconducting strings.   

After a rescaling $ \rho \longrightarrow \sqrt{\lambda_1} \rho$, $ 
z\longrightarrow \lambda_2 z $ the rescaled energy may be written 
as \be E={1 \over {\lambda_1 \lambda_2}} I_1 + \lambda_2 I_2 
+{\lambda_1 \over \lambda_2} I_3 + \lambda_1 \lambda_2 I_4 \ee 

Derrick's theorem\cite{d64} can be satisfied and collapse in any 
direction can be evaded due to the time 
dependence\cite{k97,akpf00}. We extremize E with respect to 
$\lambda_1, \lambda_2$ and set $\lambda_1\!=\!\lambda_2=1$ to obtain the 
following virial conditions \ba I_3 + I_4 &=& I_1 \label{virial1} 
\\ I_2 + I_4 &=& I_1 + I_3 \label{virial2} \ea 

In order to check the validity of these conditions numerically we 
must first solve the ode which $f$ obeys. This is of the form 
\be
f'' +{1\over \rho} f' + (\omega^2 -(2\pi N)^2/L^2) f -U'(f) = 0 
\label{fode} \ee 
with boundary conditions $ f(\infty)=0$ and  ${df\over d\rho}(0)= 
0$. Equation (\ref{fode}) is identical with the corresponding 
equation for 2D Qballs\cite{akpf00} (see ansatz (\ref{stranz})) 
with the replacement of $\omega^2$ by \be \omega^2 -{{(2\pi 
N)^2}\over {L^2}} \equiv \omega'^2\ee Solutions of (\ref{fode}) 
for various $\omega'$ and $U(f) = {1 \over 2}f^2 - {1 \over 3} f^3 
+ {B \over 4}f^4$ with $B=4/9$ were obtained in Ref. 
\cite{akpf00}. Now it is easy to see that the first virial 
condition (\ref{virial1}) may be written as 
\be
\omega'^2 \int d\rho \; \rho \; f^2 = 2 \int d\rho \; \rho U(f) 
\label{virnew} \ee This is exactly the virial theorem for 2D 
Qballs\cite{akpf00} (infinite Q strings) with $N=0$ and field 
ansatz given by (\ref{stranz}) with $\omega$ replaced by 
$\omega'$. The validity of such a virial condition has been 
verified in Ref. \cite{akpf00}. This therefore is an effective 
verification of our first virial condition (\ref{virial1}). 

The second virial condition (\ref{virial2}) can be written (using the first virial (\ref{virial1})) as
\be
2 I_3 = I_2
\ee 
which implies that
\be
{{2\pi N^2} \over L^2} = {{\int d\rho \; \rho f'^2}\over {\int 
d\rho \; \rho f^2}} \label{vir2} \ee This can be viewed as a 
relation determining the value of $L$ required for balancing the 
tension.   

These virial conditions can be used to lead to a determination of 
the energy  as \be E=2(I_1+I_3)  \ee In the thin wall limit where 
$2 \pi \int d \rho \rho f^2 = A f_0^2$ ($A$ is the surface of a 
cross section of the Q ring) this may be written as \be E \simeq 
{Q^2 \over {2 L A f_0^2}} +{{(2\pi N)^2 A f_0^2}\over {2 L}} 
\label{etwa} \ee  and can be minimized with respect to $f_0^2$. 
The value of $f_0$ that minimizes the energy in the thin wall 
approximation is 
\be
f_0=\sqrt{Q\over {2\pi N A}} \ee Substituting this value back on 
the expression (\ref{etwa}) for the energy we obtain \be E={{2 \pi 
N Q}\over L} \ee This is consistent with the corresponding 
relation for spherical Q balls which in the thin wall 
approximation lead to a linear increase of the energy with $Q$. 

The above virial conditions demonstrate the persistance of the Q 
ring configuration towards shrinking or expansion in the two 
periodic directions of the Q ring torus for large radius. In order 
to study Q rings of any size we must perform an energy 
minimization in 3D and subsequently a numerical simulation of the 
time evolution. It can be performed for any potential $U(\Phi)$ of 
a polynomial or logarithmic form that admits Q balls. In what 
follows we adopt the cubic potential \be \label{potential} U(\phi) 
= {1 \over 2}|\Phi|^2 - {1 \over 3} |\Phi|^3 + {B \over 4}|\Phi|^4 
\ee The ansatz that captures the above mentioned properties of the 
Q ring is \be \label{eq:ansatz} \Phi = f(\rho,z)\; e^{i [\omega t 
+ N\phi]} \ee where the center of the coordinate system now is in 
the center of the torus that describes the Q ring and the ansatz 
is valid for {\it any} radius of the Q ring. 

The energy of this configuration is \ba  E  &=& {1 \over 2} {Q^2 
\over \int f^2 dV} 
  + {1 \over 2} \int \left[ \left({\partial f \over \partial \rho} \right)^2
       + {N^2 f^2 \over \rho^2} \right]\;dV \nn \\
  &+& {1 \over 2} \int 
       \left[ \left({\partial f \over \partial z} \right)^2 \right]\;dV 
       + \int U(f)\,dV  \label{eq:energy2}
  \ea The field equation for $\Phi$ is 
\begin{equation}
\label{eq:equation} \ddot{\Phi} - \Delta \Phi + \Phi - |\Phi| \Phi 
+ B |\Phi|^2 \Phi = 0 
\end{equation} Substituting the ansatz (\ref{eq:ansatz}) we find that $f(\rho,z)$ 
should satisfy 
\begin{equation}
\label{eq:ansatzequation} 
 {\partial^2 f \over \partial \rho^2} 
  + {1 \over \rho}\,{\partial f \over \partial \rho}
  - {N^2 f \over \rho^2} + {\partial^2 f \over \partial z^2}
  + (\omega^2-1) f + f^2 - B f^3 = 0
\end{equation} 
In order to solve this equation we minimize the energy 
(\ref{eq:energy2}) at fixed Q using a relaxation algorithm. In the 
algorithm, we have used the initial ansatz: 
\begin{equation} 
f(\rho,z) \sim  e^{(\rho-\rho_0)+z^2}             
\end{equation} where $\rho_0$ is a fixed initial radius. 
The energy minimization resulted to a non-trivial configuration 
$f(\rho,z)$ for a given set of parameters $B, N, Q$ in the 
expression for the energy. We then used 
\begin{equation}
\label{omegdef}
\omega = {Q \over \int f^2\, dV}
\end{equation}
to calculate $\omega$ and constructed the full Q ring configuration 
using (\ref{eq:ansatz}). As a further test to the stability of the 
solution and the energy minimization algorithm we fed the 
resulting field configuration as initial condition to a leapfrog 
algorithm simulation the full field evolution in 3D by solving 
equation  (\ref{eq:equation}) in a $81^3$ lattice with reflective
boundary conditions where the second spatial derivative is set to
zero on the boundary. The relaxed  configurations with B=4/9, n=1
and various Q's were evolved for about 50 internal Q ring periods 
$T=2\pi /\omega$.
\begin{figure}[htbp] 
  \begin{center}
    \psfig{file=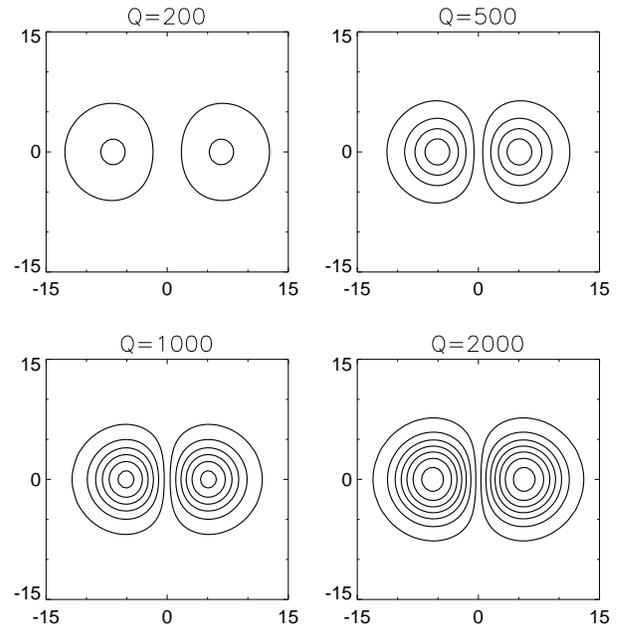,width=8truecm,bbllx=60bp,bblly=200bp,bburx=520bp,bbury=710bp}
    \caption{Charge density contour plots of x-z plane cuts for evolved 
3D profiles of relaxed Q rings ($N=1$). No significant change of 
the configurations was observed during the full 3D evolution of 
about $50$ internal rotation periods for any value of Q.} 
    \label{fig:1}
  \end{center}
\end{figure} 
\vspace{-20pt}
The contour plots of the final frames of the simulations on the 
$x-z$ plane are shown in Fig. 1. It was verified that the Q ring 
configurations evolve with practically no distortion and are 
metastable despite the long evolution. Subsequent evolution of the 
configurations shown in Fig. 1 involved non-symmetric fluctuations 
emerging from the cubic boundaries. These finite size fluctuations 
were found to lead to a break up and eventual decay of the Q ring 
to one or more Q balls (depending on the type of fluctuations) 
after more than $100$ internal rotation periods. Thus a Q ring is 
a metastable as opposed to a stable configuration. 

We have repeated the same numerical experiment with  $f(\rho,z)$ 
corresponding to a ring of large radius  with $N=0$ in the ansatz 
(\ref{eq:ansatz}) used as initial condition in the evolution 
algorithm. The result of the evolution is shown in the frames of 
Fig. 2 which shows a collapsing Q ring. It collapses due to the 
lack of pressure support ($N = 0$) within less than $10$ internal 
rotation periods in contrast to the metastable cases of Fig. 1. 
The unstable collapsing ring produces a pair of Q balls that 
propagate in opposite directions along the z-axis. This unstable 
loop is similar to the one previously seen in Q ball simulations 
of scattering in 3D \cite{bs00,lpwww}. 
\begin{figure}[htbp] 
  \begin{center}
    \psfig{file=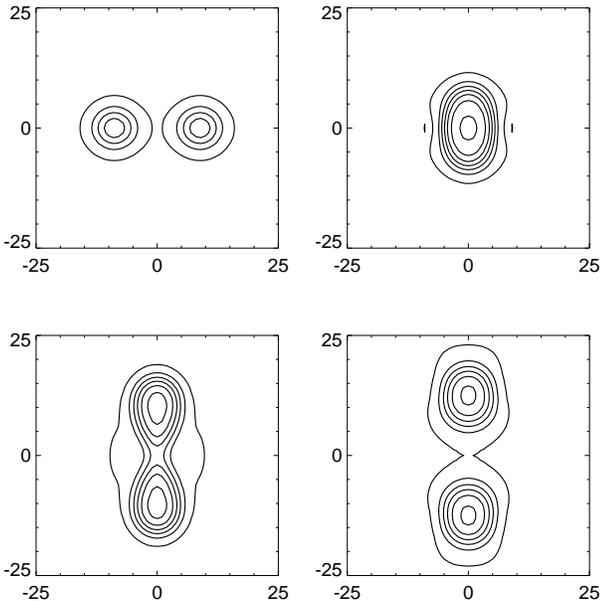,width=8truecm,bbllx=60bp,bblly=200bp,bburx=520bp,bbury=710bp}
    \caption{The 3D evolution (x-z plane cut) of an unstable Q loop with winding 
    $N=0$ significantly less time than the simulations of Fig.1 
    (about 10 internal field rotation periods). 
    The ring rapidly collapses and forms a pair of Q balls propagating
    along the z axis. Cylindrical symmetry is implied.}
    \label{fig:2}
  \end{center}
\end{figure} 
\vspace{-20pt}
The metastable Q ring configuration we have discovered is the simplest 
metastable ring-like defect known so far. Previous attempts to 
construct metastable ring-like configurations were based on pure 
topological arguments (Hopf maps) and required gauge fields to 
evade Derrick's theorem due to their static 
nature\cite{Faddeev:1997zj,Perivolaropoulos:2000gn}. Metastable 
moving non-relativistic field configurations have also been 
constructed\cite{pap}. These attempts resulted in complicated 
models that were difficult to study analytically or even 
numerically. Q rings require only a single complex scalar field 
and they appear in all theories that admit stable Q balls 
including the minimal supersymmetric standard model (MSSM). The 
simplicity of the theory despite the non-trivial geometry of the 
field configuration is due to the combination of topological with 
non-topological charges that combine to secure metastability 
without added field complications. 

The derivation of metastability of this configuration opens up 
several interesting issues that deserve detailed investigation. 
Here we outline some of these issues. \begin{itemize} \item {\bf 
Formation of Q Rings:} Q rings can form in principle by variations 
of the Kibble mechanism, the Affleck-Dine mechanism\cite{kk99} or 
by collision of Q balls with non-trivial relative phases. The 
effectiveness of these mechanisms for Q ring formation needs 
careful numerical and analytical study. 
\item {\bf Current Quenching:} The increase of the winding $N$ 
which implies increase of the current along the Q ring can not be 
arbitrary. In a way similar to superconducting string loops there 
is a maximum current and winding $N_{crit}$ beyond which the 
gradient of the phase becomes high enough to favor a zero value of 
the field inside the ring. The investigation of the dependence of 
$N_{crit}$ on Q and other parameters of the theory is a 
interesting issue. 
\item {\bf Cylindrical 
Walls:} A stabilizing phase can also be introduced in closed 
cylindrical Q walls whose tension can thus be balanced by the 
pressure of the winding phase. What are the properties of these 
objects which are essentially a simpler version of the Q rings 
discussed here? 
\end{itemize}
These and other interesting issues emerge as a consequence of our 
results and await further investigation.

\widetext 

\end{document}